\documentclass[twocolumn,3p]{elsarticle}
\usepackage{graphicx}
\usepackage{amssymb}
\usepackage[cmex10]{amsmath}
\usepackage{array}
\usepackage{listings}
\usepackage{amsfonts}
\usepackage{color}
\usepackage{algorithmic}
\usepackage{algorithm}
\usepackage{multirow}
\usepackage[caption=false]{subfig}
\usepackage{url}
\usepackage{float}
\usepackage{wrapfig}

\begin{document}

\journal{Journal of Parallel and Distributed Computing}

\title{Block-Relaxation Methods for 3D Constant-Coefficient Stencils \\ on 
GPUs and Multicore CPUs}

\author[mrs]{Manuel Birke}
\ead{manuel.birke@icloud.com}

\author[qbp]{Bobby Philip\corref{cor1}}
\ead{philipb@ornl.gov}

\author[mrs]{Zhen Wang}
\ead{wangz@ornl.gov}

\author[qbp]{Mark Berrill}
\ead{berrillma@ornl.gov}

\address[mrs]{Oak Ridge Leadership Computing Facility, Oak Ridge National 
Laboratory, Oak Ridge, TN, 37831}

\address[qbp]{Computational Engineering and Energy Sciences Group, Oak Ridge 
National Laboratory, Oak Ridge, TN, 37831}

\cortext[cor1]{Corresponding author}

\begin{abstract}
Block iterative methods are extremely important as smoothers for multigrid methods, 
as preconditioners for Krylov methods, and as solvers for diagonally dominant linear 
systems. Developing robust and efficient smoother algorithms suitable for current and 
evolving GPU and multicore CPU systems is a significant challenge.
We address this issue in the case of constant-coefficient stencils 
arising in the solution of elliptic partial differential equations on structured 
3D uniform and adaptively refined grids. Robust, highly parallel implementations 
of block Jacobi and chaotic block Gauss-Seidel algorithms with exact inversion 
of the blocks are developed using different parallelization techniques. 
Experimental results for NVIDIA Fermi/Kepler GPUs and AMD multicore systems are 
presented.
\end{abstract}

\begin{keyword}
Parallel algorithms, Graphics processors, Smoothing, Multigrid and multilevel methods, Multicore
\end{keyword}

\maketitle

\section{Introduction}\label{sec:introduction}

Iterative methods such as Jacobi, Gauss-Seidel, Successive Over-Relaxation (SOR), and their variants are 
extremely important as smoothers for multigrid methods \cite{trottenberg2001multigrid,Briggs},
preconditioners for Krylov methods \cite{Saad2003}, and as solvers for 
diagonally dominant linear systems. Their widespread use in 
iterative solution methods for linear systems has led to significant effort being devoted to optimizing their performance on modern GPU and multicore
systems. Much of the attention has been focused on point-wise versions of these 
methods to exploit parallelism \cite{DBLP:KowarschikCR04}.

However, block versions of these methods, in particular line and plane smoothers, are extremely important as key 
components of robust geometric multigrid methods \cite{Schaffer1998, Dendy1982} 
and multilevel iterative methods (such as the Fast Adaptive Composite-Grid (FAC)
and Multi-Level Adaptive Technique (MLAT) methods \cite{McCormickThomas86, Brandt1977}) on adaptively refined grids. 
Early work by Shortley and Weller \cite{shortley:334}
and Parter \cite{Parter61} has  shown that even for isotropic diffusion problems 
block Jacobi and Gauss-Seidel can have advantages. More recently, 
Philip and Chartier demonstrated how to automatically construct block iterative 
methods for fairly general linear systems based on algebraic measures of 
coupling \cite{Philip2012}. Furthermore, block iterative methods have the potential to increase
local computation and decrease communication on next generation parallel systems where communication
increasingly dominates costs.

\begin{figure*}[!t]
\centerline{
\subfloat[]{\label{fig:domain}\includegraphics[scale=1.2]{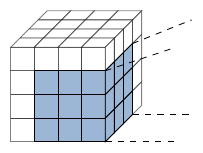}}
\subfloat[]{\label{fig:cells}\includegraphics[scale=1.2]{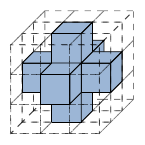}} 
\subfloat[]{\label{fig:stencil}\includegraphics[scale=1.2]{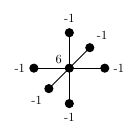}}
\subfloat[]{\label{fig:patches}\includegraphics[scale=1.2]{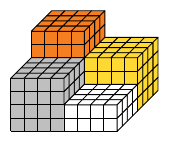}}
}
\caption{In Figure \ref{fig:domain} a patch is shown with an extraction of cells (shaded). 
The 7-point stencil in Figure \ref{fig:stencil} arises from the shaded cells in Figure \ref{fig:cells}. 
A domain composed of 4 patches is shown in Figure \ref{fig:patches}.}
\label{fig:gridStencil}
\end{figure*}

In this context, recent work focused on GPU implementations includes 
the 2D block based smoother of Feng et al. \cite{Feng:2008:MGT:1509456.1509599} 
and the 1D block-asynchronous smoother for multigrid methods by 
Anzt et al. \cite{anzt2011block}. However, both works use Jacobi iterations within each 
block, rendering them closer to two-level Jacobi methods or Jacobi-Gauss-Seidel 
methods. These can be considered as variants on the work presented in 
Venkatasubramanian et al. \cite{Venkatasubramanian:2009}. Recent work in the context of multicore CPUs includes that by Adams et al. 
\cite{Adams2003593,Adams:2001:DMU:582034.582038} on block 
Gauss-Seidel algorithms. 

In this paper we focus on developing efficient block-iterative Jacobi and 
chaotic Gauss-Seidel methods on current and evolving GPU and multicore architectures. We will demonstrate simple, efficient and general
block smoothing algorithms with exact inversion of the blocks for 3D constant-coefficient elliptic problems.
Such problems are of importance in a wide variety of scientific computations.
We will consider only single node GPU and multicore algorithms in this paper
with future work extending the results of this paper to distributed systems.
The processing power and memory capacity of single node systems
enable fairly large simulations particularly when Adaptive Mesh Refinement (AMR) is used
where the memory and compute requirements drop significantly
(approximately 10\% of uniform grid computations \cite{Philip2008,Pernice2005} in our experience),
making the results of this paper relevant in that context.

\subsection{Model Problem}

We will focus on the model problem for the 3D Poisson equation 
\begin{eqnarray}\label{eq:poisson}
-\nabla^{2}u\left(\mathbf{x} \right)&=&f(\mathbf{x} ), \; \; \mathbf{x} \equiv \left(x,y,z\right) \in \Omega, \nonumber \\
u\left(\mathbf{x}\right)&=& 0, \; \; \mathbf{x} \in \partial \Omega,
\end{eqnarray}
\noindent for simplicity, though the presented methods will be applicable to 
any constant-coefficient elliptic system such as the recent work by 
Guy et al. \cite{Guy2012} on block smoothers for Stokes problems.
Here $\nabla^{2}$ is the Laplacian operator, $f$ is a source and $u$ is the  
solution to Equation (\ref{eq:poisson}) on a cubic 
domain $\Omega \equiv \left[0,1\right]^{3} \subset \mathbb{R}^{3}$ 
with Dirichlet boundary conditions on the boundary $\partial \Omega$. We set 
$f\left(x\right)=0$ for simplicity because our focus is on the smoothing 
algorithm.

\begin{figure}[!t]
\centerline{
\subfloat[]{\label{fig:amr}\includegraphics[scale=1.0]{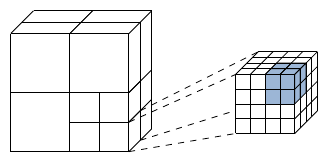}}
\subfloat[]{\label{fig:ghosts}\includegraphics[scale=1.0]{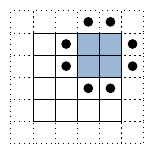}}
}
\caption{ (a) Left: 3D AMR patch hierarchy with outlines of 8 patches 
on level 0 and 8 patches on level 1. Right: patch with shaded cells 
corresponding to a geometric block. (b) 2D patch slice
with ghost and neighbor cells for a block (shaded).}
\label{fig:amrGrid}
\end{figure}

It is assumed that the reader is familiar with discretization methods and only a brief description is provided to establish notation. The methods presented are not tied to any particular discretization, though for concreteness we use a cell-centered finite volume (FVM) discretization with variable unknowns located at cell centers. FVM discretization methods for single logically rectangular domains begin by partitioning the continuous domain $\Omega$ into a set of discrete cells that together form a regular patch with $n_{x}$, 
$n_{y}$, and $n_{z}$ cells in the $x$-, $y$-, and $z$-directions respectively as in Figure \ref{fig:domain}. Each cell volume is then uniquely indexed by a triple $(i,j,k)$ specifying location in space: $\left[i \cdot h_x,\left(i+1\right) \cdot h_x\right] \times 
\left[j \cdot h_y,\left(j+1\right) \cdot h_y\right] \times 
\left[k \cdot h_z,\left(k+1\right) \cdot h_z\right]$ with $0 \leq i \leq n_{x}-1$, $0 \leq j \leq n_{y}-1$, $0 \leq k \leq n_{z}-1$. Here $h_x, h_y,$ and $h_z$ represent the cell widths in each direction. To simplify notation going forward we assume $h_x = h_y = h_z = h$.
A cell-centered finite-volume approximation of equation (\ref{eq:poisson}) then leads to the following system of equations for each cell:
\begin{eqnarray}\label{eq:stencil3d7}
6u_{i,j,k} - u_{i-1,j,k} - u_{i+1,j,k} - u_{i,j-1,k} \nonumber\\ - u_{i,j+1,k} - u_{i,j,k-1} - u_{i,j,k+1} = h^2 f_{i,j,k},
\end{eqnarray}
where $u_{i,j,k}$ represents an approximation to $u$ in cell $\left(i,j,k\right)$.
Since the coefficients in Equation (\ref{eq:stencil3d7}) have no spatial dependence 
a stencil representation as depicted in Figure \ref{fig:stencil} can 
be used to represent both the coefficient and connectivity information for 
each cell in a patch. Cells on patch boundaries have the same stencil as 
interior cells, since we assume that each patch has a layer of `ghost' 
cells around it (Figure \ref{fig:ghosts}) to which boundary conditions 
are extrapolated. The methodology outlined extends with minor modifications 
to discretizing a collection of non-overlapping logically rectangular patch 
domains as in Figure \ref{fig:patches}. Interior ghost cells at patch 
interfaces are then used to ensure consistency across patches.

Many simulations of physical phenomena often need to resolve extremely fine-scale features localized in space and/or time \cite{Philip2008,Pernice2005}. Discretizing the entire physical domain at the required fine-scale resolution is then often impractical and not necessary. Instead AMR can be used to increase the local resolution only where required. Given a collection of patches that cover the domain $\Omega_1 \equiv \Omega$ at some coarse resolution $h_1 = h$, structured AMR techniques identify a local subdomain $\Omega_2$ where finer resolution is required ($\Omega_2$ may consist of disjoint subdomains).  A collection of patches with the same finer resolution $h_2$ are then introduced to cover $\Omega_2$. Together, they form a refinement level covering the subdomain $\Omega_2$. A typical choice of $h_2$ is $h_1/2$. Repeating this process leads to a set of increasingly finer nested refinement levels, each consisting of a collection of logically rectangular patches. Patches in this hierarchy are dynamically created and destroyed as the simulation progresses depending on local resolution requirements. Large-scale parallel structured AMR calculations can employ thousands of patches on each refinement level, with each processor owning multiple patches from potentially different refinement levels. Methods such as FAC \cite{McCormickThomas86} and MLAT \cite{Brandt1977} smooth on these patches during the solution process. 
For the purposes of block smoothing, patches in the refinement hierarchy are typically 
too large to be processed efficiently. Instead patches are decomposed 
into smaller geometric blocks. 
For example, Figure \ref{fig:amr} shows a $2\times2\times2$ block (shaded) consisting 
of 8 cells in a $4\times4\times4$-cell patch.
Thus, the patch of size $4^{3}$ cells in Figure 
\ref{fig:amr} is decomposed into 8 blocks of size $2^{3}$. Stencil operations 
for each block in the patch have dependencies on neighbor and ghost cell data as 
shown in Figure \ref{fig:ghosts}.
As mentioned, the block relaxation methods we describe below
are meant to be used as components of multigrid or multilevel methods
consisting of multiple structured patches over a domain or a hierarchy of sub-domains in the case of AMR.

The remainder of the paper is organized as follows. Section \ref{sec:algorithm} describes block-relaxation methods and algorithm choices based on GPU and multicore CPU architectures. Section \ref{sec:blockDomain} discusses different parallelization strategies 
and considerations in the choice of block sizes. We state experimental results using state-of-the-art hardware including 
the NVIDIA Kepler K20X GPU and a CRAY XE6 compute node in Section \ref{sec:results}. 
The last section presents conclusions and an overview 
of possible future research directions.

\section{Block Smoothing Algorithms}\label{sec:algorithm}

Discretizing equation (\ref{eq:poisson}) on a patch or a collection of patches leads to a 
linear system of equations, one equation per grid cell, of the form of equation (\ref{eq:stencil3d7}). The resulting linear system of equations  
can be written in matrix form as
\begin{eqnarray}\label{eq:system} 
A\mathbf{u} & = & \mathbf{f}
\end{eqnarray}
\noindent with $A \in \mathbb{R}^{m \times m}$, and $\mathbf{u}, \mathbf{f} \in \mathbb{R}^{m}$ with a suitable mapping from the matrix ordering to the $(i,j,k)$ ordering on patches. Let $A$ be partitioned into a set of submatrices as shown below.
\begin{eqnarray}\label{sm:blocka}
A  = \left[ \begin{array}{cccc}
                A_{11} & A_{12} & \cdots & A_{1s} \\
                A_{21} & A_{22} & \cdots & A_{2s} \\
                \cdots   & \cdots  & \cdots & \cdots \\
                A_{s1} & \cdots  & A_{s,s-1} & A_{ss}
                \end{array}
         \right]
\end{eqnarray}
where $A_{ij},\,1\leq i,j \leq s$ are now $s^2$ matrix subblocks of
size $q_i\times q_j$ with $\sum_{1}^s q_i = m$. The partitioning of $A$ into subblocks could be based on mapping to/from geometric blocks of a structured grid as described earlier or on other considerations such as anisotropic features in the PDE, or algebraic strength of coupling measures between variables.  It is assumed that the diagonal blocks, $A_{ii}, 1\leq i\leq s$, are invertible. For the purposes of this paper it is sufficient to consider the matrix blocks as arising from a
lexicographical ordering of geometric blocks within each patch. Furthermore, from this point on we will assume that all matrix blocks are of the same size. A block stationary iterative method can now be defined by a splitting, $A=M-N$, where $M$ is invertible, which leads to the stationary
iteration
\begin{eqnarray}\label{sm:blockgsf}
\mathbf{u}^{k+1}  = M^{-1}N\mathbf{u}^k+M^{-1}\mathbf{f},
\label{blockiteration1}
\end{eqnarray}
where $k$ is the iteration number.
The iteration given above converges if and only if $\rho(M^{-1}N)<1$, where $\rho$ denotes the spectral radius operator. An equivalent formulation of (\ref{blockiteration1}) is:
\begin{eqnarray}\label{sm:blockgs}
\mathbf{u}^{k+1}  = \mathbf{u}^k+M^{-1}\mathbf{r}^k
\label{blockiteration2}
\end{eqnarray}
for residual $\mathbf{r}^k=\mathbf{f}-A\mathbf{u}^k$, which will be the form used in this paper.
An example of a block iteration is the block Gauss-Seidel method defined by the 
splitting
\begin{eqnarray}\label{sm:blockm}
M = \left[ \begin{array}{cccc}
                A_{11} & 0 & \cdots & 0 \\
                A_{21} & A_{22} & \cdots & 0 \\
                \cdots   & \cdots  & \cdots & \cdots \\
                A_{s1} & \cdots  & A_{s,s-1} & A_{ss}
                \end{array}
         \right]
\end{eqnarray}
and $N=M-A$. When $s=m$, the subblocks are of size one and the
iteration reduces to the standard lexicographic Gauss-Seidel
iteration. A block Jacobi algorithm is obtained if $A_{ij}=0$ for $i>j$ also in equation (\ref{sm:blockm}). The numerical results presented in this paper were all
performed with block (chaotic) Gauss-Seidel and damped block Jacobi iterations, though once the 
block partitioning is defined we are free to choose any suitable block-iterative process.

In general, it is difficult to provide theoretical results that guarantee that a given block-iterative method 
formed will converge or that a particular choice of blocks will lead to a faster rate of convergence as opposed to an alternative choice of blocks. The interested reader is referred to Varga \cite{Varga99} for theoretical results on general block iterative methods and to Parter \cite{Parter61} for some early work on block iterative methods for elliptic equations.

The block-smoothing iteration derived from Equation (\ref{sm:blockgs}) employs 
a direct solve to compute the inverse $\hat{A}_{b}:=A_{ii}^{-1}$ for the diagonal block $A_{b}:=A_{ii}$ 
with $b,i=1,\ldots,s$. 
The cost of computing this inverse for a block, and restrictions that apply in 
the case that the patch size is not a multiple of the block size, are
described in Section \ref{sec:binpact}. There it will be shown that it is sufficient to assume that only one 
inversion is necessary for a patch and this inversion has already been performed 
before the smoothing iterations are executed.

Updating the neighbor cells of a spatial block as in Figure \ref{fig:ghosts} 
can be performed in different ways. 
Our implementation of the block smoothing algorithm employs either a 
classic Jacobi update or a chaotic block Gauss-Seidel update scheme between 
the blocks. This \emph{chaotic relaxation} was first proposed by Chazan 
and Miranker \cite{Chazan1969199}. Some current research on chaotic relaxation 
using shared and distributed memory systems can be found under 
\emph{asynchronous iteration} in 
\cite{Frommer:2000:AI:363882.363901,Baudet:1978:AIM:322063.322067,Strikwerda2002125}.
Baudet \cite{Baudet:1978:AIM:322063.322067} defines 
the term \emph{chaotic relaxation scheme} to describe a 
purely or totally asynchronous method, which accurately describes our chaotic block 
Gauss-Seidel implementation. Bertsekas et al. state in \cite{bertsekas1989parallel} 
a general convergence theorem for totally asynchronous algorithms in the case 
of fixed-point problems. A modified convergence theorem based on Chazan and Miranker
with constraints on global memory and communication is presented by 
Strikwerda \cite{Strikwerda199715}. Blathras et al. state a timing model and 
stopping criteria for block asynchronous iterative methods in 
\cite{Blathras99timingmodels}. A basic outline of the chaotic block Gauss-Seidel 
relaxation for multiple patches either on an AMR refinement level or a multigrid level 
is shown in Algorithm \ref{alg:smoothBase}, assuming that the inner most loop will 
be executed asynchronously.
\begin{algorithm}
\caption{Chaotic block Gauss-Seidel relaxation algorithm.\label{alg:smoothBase}}
\begin{algorithmic}[lines]
\FOR{each step $k$}
\FOR{each patch $p$}
\FOR{each block $b$}
\STATE $u^{k+1}_{b} = u^{k}_{b} + \hat{A}_{b} \left(f_{b} - A_{b}u^{k}_{b} \right)$ 
\ENDFOR
\ENDFOR
\ENDFOR
\end{algorithmic}
\end{algorithm}
In the block Jacobi update scheme a second vector $v$ stores the new values 
of a patch during the smoothing step. At the end of the iteration the solution vector 
$u$ for a patch swaps with $v$. This  can be done in a very efficient way by 
using different pointers to the corresponding memory areas so that no 
additional memory copy is necessary. The chaotic block Gauss-Seidel update 
scheme does not need an extra vector. Advancing the solution can be done 
immediately. 

No synchronization between the matrix blocks is necessary while processing the 
blocks within a smoothing step when using block Jacobi iteration or 
chaotic block Gauss-Seidel relaxation. Therefore the blocks can be processed 
independently without any communication. The ghost cells between patches are 
updated in a Jacobi fashion at the end of a complete smoothing step over all 
patches. To prevent confusion about blocks on the GPU and blocks in the 
physical domain, a block in the domain 
is called a \emph{spatial block} and a block on the GPU is called 
a \emph{thread block}. 
The structure of AMR leads to the possibility to parallelize only over patches 
(\emph{patch parallel}) or only over spatial blocks (\emph{block parallel}) leading to \emph{single-level parallelism}.
Parallelizing over patches and then over spatial blocks leads to \emph{two-level} or \emph{nested parallelism}. 

A GPU is well suited to exploit the fine-grained parallelism typically exposed by single-level 
methods. It is also possible to exploit the parallel architecture of a GPU 
using the two-level parallelism by processing several patches and spatial blocks concurrently. 
For two-level parallelism the number of patches that can be processed in 
parallel depends on the number of spatial blocks that can be scheduled and executed 
independently on the GPU. This then depends on the block size, the hardware 
architecture of the GPU, and the number of spatial blocks that are executed per patch.

\begin{figure*}[!t]
\centerline{
\subfloat[]{\label{fig:patchset}\includegraphics[scale=1.2]{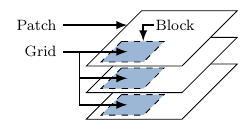}}
\subfloat[]{\label{fig:blockset}\includegraphics[scale=1.2]{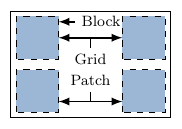}} 
\qquad
\subfloat[]{\label{fig:largetb}\includegraphics[scale=1.2]{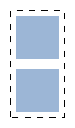}} 
\subfloat[]{\label{fig:sweeptb}\includegraphics[scale=1.2]{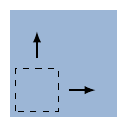}} 
\subfloat[]{\label{fig:sametb}\includegraphics[scale=1.2]{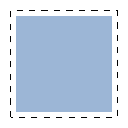}} 
}
\caption{Examples demonstrating how to map patches to a GPU grid and how to map 
a spatial block to a thread block. Figure \ref{fig:patchset} shows several 
patches that are mapped to one GPU grid and \ref{fig:blockset} shows one 
patched mapped to one GPU grid. In Figure \ref{fig:largetb} the thread block 
(dashed) is larger than the spatial block (shaded), in \ref{fig:sweeptb} 
a thread block is smaller than the spatial block and in \ref{fig:sametb} a 
thread block has the same size as a spatial block.}
\label{fig:mappPatch}
\end{figure*}

On multicore CPUs scalable performance is typically achieved using coarse-grained parallelism. 
The work per thread must be large enough to amortize the overhead of scheduling the work. 
The block-parallel algorithm
in which the spatial blocks are processed in parallel has a potential bottleneck if 
the work per thread is low for small block sizes. Using the patch parallel method increases 
the workload per thread so that this bottleneck is avoided.
Furthermore, two-level parallelism can improve the performance on a CPU beyond 
ordinary single-level parallelization \cite{Blikberg:2005:LBO:1131934.1131937}
especially in comparison to block-level parallelism,
which is also demonstrated in Section \ref{sec:results}.

\subsection{Implementation for CUDA}\label{sec:implementCuda}

The implementation of block-relaxation algorithms on GPUs is done using 
the NVIDIA Compute Unified Device Architecture (CUDA). A short introduction 
into the technology and programming of CUDA can be found in 
\cite{Nickolls:Cuda,Kirk:Cuda}.  
Threads in CUDA are grouped together into thread blocks and are executed 
in a Single Instruction Multiple Threads (SIMT) fashion of 32 
threads called a \emph{warp}. Thread blocks are 
organized into grids and each CUDA kernel launches one grid of several thread blocks. 

The decomposition of a patch in spatial blocks shares similarities with the CUDA 
architecture. But mapping a spatial 
block to a thread block or a patch to a grid on the GPU is nontrivial 
in the case of block smoothing. 
The mapping has an impact on the level of parallelism that can be 
achieved using different spatial block and patch sizes. 
We suggest the following two strategies for processing patches on the GPU:

\emph{Strategy 1:} One thread block processes all spatial blocks within a patch 
as in Figure \ref{fig:patchset}. The thread block \emph{sweeps} through the 
patch. In this case a GPU grid maps different thread blocks to different
patches. Spatial blocks in different patches are processed 
in parallel resulting in two-level parallelism.

\emph{Strategy 2:} One patch is processed by several thread blocks on the GPU 
as shown in Figure \ref{fig:blockset}. 
Here single-level parallelism exists if the patch size is large enough that the 
maximum number of scheduled thread blocks on the GPU is reached. In this case only 
one patch can be processed at a time on the GPU. If the patch size is small 
enough that multiple patches can be processed in parallel, then two-level 
parallelism is possible. 

In this paper we explore Strategy 2 for the presented 
implementations of the block smoothing relaxation. Strategy 1 will be 
explored in future research projects. In addition to the patch mapping, the spatial blocks must be mapped to the 
thread blocks on the GPU.
A common method for spatial blocking is overlapping axis-aligned 
3D blocking \cite{Nguyen:2010:BOS:1884643.1884658}, in which a spatial block 
of dimension 
$b_{x} \times b_{y} \times b_{z}$ with neighbor cells 
$\mathcal{R}$ is loaded into on-chip memory, leading to some 
redundant memory transfers for the neighbor cells. 3D blocking can be optimized 
by a 2.5D sliced blocking technique such as 
\cite{Micikevicius:2009:FDC:1513895.1513905} or 
\cite{Nguyen:2010:BOS:1884643.1884658}. In 2.5D slicing, a block is 
decomposed into slices of cells. These slices are processed sequentially in 
an alternative blocking that preserves some cells in the on-chip cache or 
even registers, reducing global memory access. 2.5D blocking is outside
the scope of this paper but will be explored in future work. We focus on 3D 
blocking of the domain. Methods of creating a mapping from the 
spatial block to the thread block are:
\begin{itemize}
\item[A.]  Thread blocks are larger than spatial blocks. 
Several spatial blocks can be processed by one thread block 
(Figure \ref{fig:largetb}). 
\item[B.] Thread blocks are smaller than spatial blocks. 
The thread blocks have to \emph{sweep} through each spatial block
(Figure \ref{fig:sweeptb}).
\item[C.] Thread blocks are the same size as spatial blocks. 
One thread block processes one block in the spatial domain
(Figure \ref{fig:sametb}).
\end{itemize} 
The 2.5D slicing is the (B) block mapping, where the thread block sweeps
through slices of a spatial block.
Mapping (C) is used for the implementation of the block smoothing 
algorithm described in this paper. This mapping differs in two major ways 
from the other mappings. 
First, it is not necessary to sweep through the spatial domain,
avoiding the introduction of additional \emph{for-loops} as would be necessary for 
mapping (B). Second, there is no need for 
a more fine-grained partitioning like in mapping (A). A fine-grained partition
means that in order to identify a corresponding spatial block for a thread 
in mapping (A) additional indexing must be performed. 

In mapping (C) each thread in a thread block $b$ is assigned a 3D 
index $\left(i,j,k\right)$ 
corresponding directly to the cell with the cell-centered spatial index. 
A thread resides in two index spaces: the 
\emph{global 3D space} that is the index in the patch and the 
\emph{local 3D space} that is the index within a thread block. 
The global 3D index calculation in CUDA without ghost cells using 
standard lexicographical ordering can be performed in the following manner:
\begin{equation}\label{eq:index3d}
\begin{split}
i = blockIdx.x \times blockDim.x + threadIdx.x \\
j = blockIdx.y \times blockDim.y + threadIdx.y \\
k = blockIdx.z \times blockDim.z + threadIdx.z.
\end{split}
\end{equation}
For the calculations with respect to the ghost cells, the thread index must 
be shifted by the ghost cell width. The local 3D index is given by the 
three \emph{threadIdx} values in CUDA.

Cells of a patch including the ghost values are stored in a 1D array. 
The right-hand side of Equation (\ref{eq:system}) needs no ghost values. 
Accessing both vectors on the GPU kernel requires different indexing. Thus a 
thread resides in additional 1D index spaces: the \emph{global 1D space} 
with and without ghost values. If the on-chip shared memory is used within 
a thread block, an additional \emph{local 1D space} is also required. The index 
calculation for the 1D space is a simple axis projection either on the 
local $blockDim$ or the global $blockDim \times gridDim$ axis. 
In the resulting block Jacobi iteration on the GPU, as shown in Algorithm 
\ref{alg:jacobiSmooth}, each thread in a thread block $b$ 
computes the new value of a cell in the physical domain. 
\begin{algorithm}
\caption{Block Jacobi algorithm on GPUs.\label{alg:jacobiSmooth}}
\begin{algorithmic}[lines]
\STATE \COMMENT{Step 1 - Thread block computes residual}
\STATE $r_{b}=f_{b}-A_{b}u_{b}$
\STATE \COMMENT{Barrier for shared memory update}
\STATE \_\_syncthreads()
\STATE \COMMENT{Step 2 - Thread block computes dgemv}
\STATE $r_{b} = \hat{A}_{b} r_{b}$
\STATE \COMMENT{Step 3 - Thread block advances solution}
\STATE $v_{b} = u_{b} + \omega r_{b}$
\STATE \COMMENT{Step 4 - Swap the vectors}
\STATE $u = v$
\end{algorithmic}
\end{algorithm}
In the first step of Algorithm \ref{alg:jacobiSmooth} the computation of the 
block residual $r_{b}$ is performed, including the 7-point stencil operation 
from Equation (\ref{eq:stencil3d7}) that each thread performs. 
The computed residual $r_{b}$ for a particular spatial block $b$ is stored in 
shared memory on the GPU because all threads in a block must access  
the residual in order to compute the matrix-vector operation in step 2. 

The second step of Algorithm \ref{alg:jacobiSmooth} consists of multiplying 
the block-diagonal inverse $\hat{A}_{b} \in \mathbb{R}^{n \times n}$ 
(where $n$ is the product of the block dimensions) with the block residual $r_{b}$. 
Each thread in a thread block performs a dot product for its particular 
cell. The matrix is stored in column-major ordering to optimize the 
memory access pattern. 
Step 3 advances the solution of a spatial block. In the case of Jacobi iteration 
the relaxation parameter $\omega = 0.8$ is used and the new block solution is 
stored in a second vector $v$. The chaotic block Gauss-Seidel updates the solution to 
$u_{b} + \omega r_{b}$ with $\omega = 1$ without any intermediate vector $v$.
Swapping the vector $v$ with $u$ in step 4 is only necessary for the 
block Jacobi iteration. 

A feature of the NVIDIA GPUs is that the GPU supports 
concurrent data transfers, transfer-execution or kernel-execution 
overlapping \cite{CUDAGuide}. Depending on the instruction stream in the 
algorithm, either concurrent data transfers, transfer-execution or kernel-execution 
overlapping is performed.

For transfer-execution overlapping, the instructions must be executed in 
\emph{depth-first} order where for all patches the instruction chain: 
memory transfer to the GPU, kernel execution and the memory transfer 
back to the host is executed. In this case the transfers from and to the GPU memory 
can overlap with kernel execution. To execute kernels 
concurrently, the sequence of instructions must be in \emph{breadth-first} 
order in which the transfer for all patches is started, then  
the kernel execution for all patches and at last the transfer back to the 
host for all patches. The overlapping of kernel executions is possible only  
when the number of blocks 
executed by one kernel does not exceed the number of blocks that can be 
scheduled concurrently on the GPU. 

\subsection{Implementation for a Multicore Architecture}

In the serial implementation of block Jacobi, 
Algorithm \ref{alg:jacobiSmoothHost} will be executed for each patch. 
The CPU implementation first computes the block residual. After this step 
the block residual is multiplied with the spatial block inverse in step 2. This 
is done using the optimized Basic Linear Algebra Subprograms 
(BLAS) Level-2 (dgemv) function. Step 3 must be 
executed separately from step 2 because accessing the vector $v$ requires a 
global 3D index computation.
Finally advancing the solution using the block Jacobi update scheme has to 
be done after all blocks are processed in step 4.

Algorithm \ref{alg:jacobiSmoothHost} is parallelized with OpenMP \cite{ARB:OpenMPSpec} 
directives by implementing the single-level and two-level 
parallelism. OpenMP is a compiler extension that 
enables portable, automated thread parallelism for multicore CPUs. Additional 
information about OpenMP programming 
can be found in \cite{Chandra:ParallelOpenMP,Chapman:UsingOpenMP}.
For the 
single-level parallelism the implementation parallelizes either the processing 
of patches or the processing of blocks. In the two-level parallelism $p$ threads 
process the patches and spawn additional $q$ threads to 
process the blocks. 

An OpenMP \emph{parallel for} loop is used to parallelize in the single-level parallelism 
over either the blocks or the patches. For the two-level parallelism an OpenMP 
\emph{parallel region} is created to parallelize over the patches and an additional region is 
created to parallelize over the blocks using the OpenMP 3.0 \emph{collapse} 
directive to collapse the loops over the 3 dimensions $x,y,z$.
\begin{algorithm}
\caption{Block Jacobi algorithm on CPUs.\label{alg:jacobiSmoothHost}}
\begin{algorithmic}[lines]
\FOR{block index in $Z$}
\FOR{block index in $Y$}
\FOR{block index in $X$}
\STATE \COMMENT{Step 1 - Compute block residual}
\STATE $r_{b}=f_{b}-A_{b}u_{b}$ 
\STATE \COMMENT{Step 2 - Compute dgemv}
\STATE $r_{b} = \hat{A}_{b}r_{b}$
\STATE \COMMENT{Step 3 - Create block solution}
\STATE $v_{b} = \omega r_{b}$
\ENDFOR
\ENDFOR
\ENDFOR
\STATE \COMMENT{Step 4 - Advance the patch solution}
\STATE $u = u + v$
\end{algorithmic}
\end{algorithm}

The chaotic block Gauss-Seidel version of Algorithm \ref{alg:jacobiSmoothHost} 
uses only the single-level block parallelism or nested parallelism to process the 
blocks asynchronously.

\section{Estimating Spatial Blocking}\label{sec:blockDomain}

\begin{figure*}[!t]
\centerline{
\subfloat[Relative error.]
{\label{fig:convError}\includegraphics[scale=1.0]{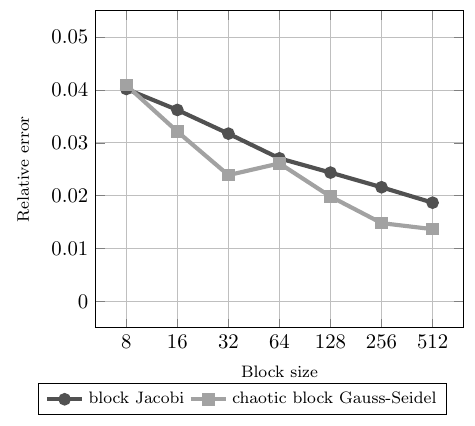}} \hfill
\subfloat[Convergence factor.]
{\label{fig:conFac}\includegraphics[scale=1.0]{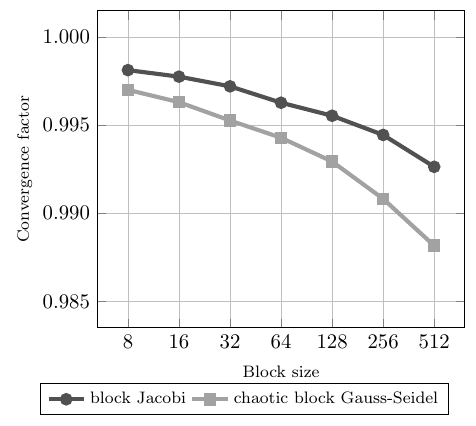}} 
}
\caption{Convergence study of block Jacobi and chaotic block Gauss-Seidel 
relaxation.}
\label{fig:convStudy}
\end{figure*}

The block size in the physical domain influences the convergence 
and smoothing properties of the algorithm. For multilevel algorithms it is the 
latter criteria that is of more importance and is also less well understood 
theoretically other than for specific cases. However, it is important that the 
algorithm converges all error modes and our numerical experiments in this 
subsection demonstrate this.

The results of numerical experiments on convergence behavior with different block sizes 
for block Jacobi and chaotic block Gauss-Seidel relaxations are presented in 
Figure \ref{fig:convStudy}. The following block sizes 
$\left(b_{x},b_{y},b_{z}\right)$ are used: 
$\left(2,2,2\right)=8$, $\left(4,2,2\right)=16$, $\left(4,4,2\right)=32$,
$\left(4,4,4\right)=64$, $\left(8,4,4\right)=128$, $\left(8,8,4\right)=256$, 
$\left(8,8,8\right)=512$. We use $\left(4,2,2\right)$ instead of 
$\left(2,4,2\right)$ or $\left(2,2,4\right)$ because  
the $x$-dimension provides the fastest memory access in our implementation. We note that
for problems with anisotropic coefficients this may not be the best choice.

For the convergence study a patch size of $64^{3}$ is used. The relative 
error estimate is computed after 3 smoothing steps with a random initial guess. In multigrid applications,
performing at most 3 smoothing steps is usually sufficient to smooth the oscillatory error modes. The 
selection of the $64^{3}$ patch size is based on the volume to surface-area 
ratio and the overall memory required to store a patch.
Considering a ghost cell width of 1, a $32^{3}$ domain with ghost cells needs 
overall $34^{3}$ cells. The memory overhead 
to store the ghosts cells is 6536 cells, around 20\% of the overall 
memory needed for $32^{3}$ cells. By using a patch size of $64^{3}$ the overhead is 
only 10\% and for $96^{3}$ only 6\%. 
Larger patches need more memory space so that a patch size 
of $96^{3}$ requires about 14 Megabytes in double precision to store $u$ and 
$f$ for all cells in the Gauss-Seidel algorithm and about 21 Megabytes in the 
Jacobi algorithm. With bigger patches the number of patches that can be processed 
within a GPU or CPU drops which impacts the balance of the patch distribution 
across different processes. Choosing patches between $64^{3}$ and $96^{3}$ is 
a trade-off between the number of patches, memory usage and load balancing.  

In Figure 
\ref{fig:convError} the relative error 
$\left\|{r_{k}}\right\|/\left\|{r_{0}}\right\|$ is 
shown for different block sizes for $k=3$. Figure \ref{fig:conFac} shows the 
asymptotic behavior of 
the convergence factor $\sqrt[k]{\left\|{r_{i+k}}\right\|/\left\|{r_{i}}\right\|}$ 
where $i,k$ defines a sliding window for a random 
initial guess and a large number of smoothing steps ($i \approx 5000$). The asymptotic convergence rate 
for block Jacobi is reached after a relatively few number of steps while in the case of chaotic block Gauss-Seidel the 
rate will jitter from step to step because of the chaotic updates but nevertheless the 
estimation is fairly accurate.

The block Jacobi and chaotic block Gauss-Seidel convergence study shows an overall 
monotonic improvement in relative error and convergence factor by increasing 
the block size for a random initial guess. 
The convergence in smoothing is necessary but does 
not reflect in general which types of modes have been smoothed and is outside the scope of this paper. However, the 
convergence behavior serves as a baseline for the performance evaluation 
with respect to the block size for the presented implementation.

\section{Experimental Results}\label{sec:results}

\begin{figure*}[!t]
\centerline{
\begin{tabular}{cc}
\subfloat[Block Jacobi block size of $4^{3}$.]
{\label{fig:vayuJacobiSpeed4}\includegraphics[scale=1.0]{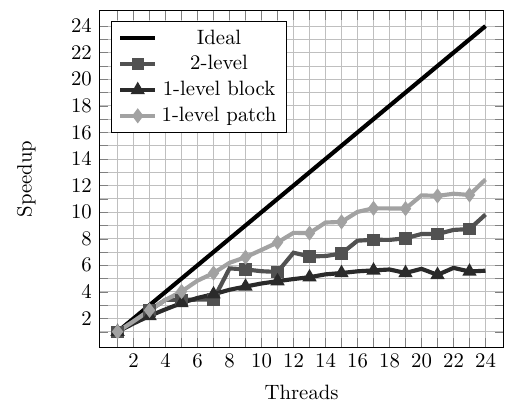}} &
\subfloat[Block Jacobi block size of $8^{3}$.]
{\label{fig:vayuJacobiSpeed8}\includegraphics[scale=1.0]{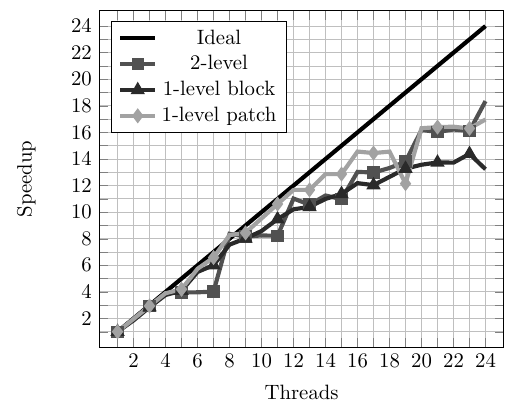}} \\
\subfloat[Chaotic block Gauss-Seidel block size of $4^{3}$.]
{\label{fig:vayuGaussSpeed4}\includegraphics[scale=1.0]{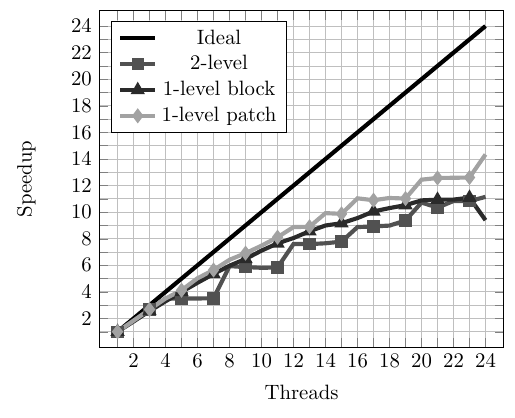}} &
\subfloat[Chaotic block Gauss-Seidel block size of $8^{3}$.]
{\label{fig:vayuGaussSpeed8}\includegraphics[scale=1.0]{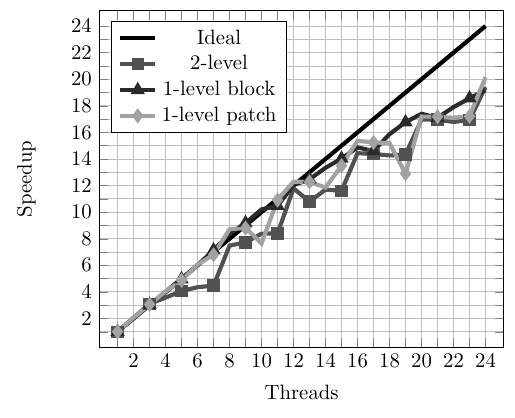}}	
\end{tabular}
}
\caption{Speedup of block Jacobi and chaotic block Gauss-Seidel relaxation using 
different levels of parallelism.}
\label{fig:speedupJacobi}
\end{figure*}

This section presents the experimental results for various GPU and multicore 
CPU implementations of our block smoothing algorithm. In Section 
\ref{sec:speedup} the speedup of a 24-core Maranello system will be presented. 
The speedup $S(p)$ of $p$ processors is defined as the fraction of 
sequential over parallel wall time on $p$ processors. Estimating $S(p)$ 
with a constant problem size and increasing processor count is called 
\emph{strong scaling}. A speedup of $S(p)=p$ is \emph{ideal}, as 
the wall time improves proportional to the number of processors that are added. Strong scaling
gives a good understanding of how well the algorithm scales with
the tested parameter set. We can derive the efficiency $E(p)$ 
of a parallel system with $p$ processors from the speedup.
The efficiency is defined by the ratio $S(p)/p$ between the 
speedup and the number of processors that reaches 
the particular speedup.

Section \ref{sec:appperf} compares the wall time of different parameter sets executing the 
block-relaxation algorithms on the GPUs and CPUs. For the benchmarks 
different hardwares including a CRAY XK7 node from the 
``Titan'' supercomputer at Oak Ridge National Laboratory are used.
Finally in Section \ref{sec:binpact}, 
we provide detailed benchmarks for 
inverting the diagonal-block matrices of equation (\ref{sm:blockm}) 
with different block sizes.
All measurements presented in this section include the update of the 
ghost cells between patches.

\subsection{Multicore Speedup}\label{sec:speedup}

A state-of-the-art Symmetric Multiprocessing (SMP) compute node equipped  
with two AMD Opteron 6168 Magny Cours CPUs is used to benchmark the  
multicore implementations. Each CPU contains 2 dies with 12 cores, and each die has 6 
shared 256 bit floating-point units. Using the two-level parallelism 4 threads 
(number of dies that the system has) are spawned 
to process the patches in parallel and overall 24 threads are processing the 
blocks in parallel. Figure \ref{fig:speedupJacobi} shows the strong-scaled 
speedup for the block Jacobi and chaotic block Gauss-Seidel algorithm processing 
96 patches with 
patch size of $64^{3}$ using two different block sizes. 
The selected block sizes are $4^{3}$ and $8^{3}$ because the experiments in Section 
\ref{sec:blockDomain} suggest a good convergence factor for $8^{3}$. 
The block size of $4^{3}$ is used to provide a contrast to the $8^{3}$ with respect 
to the cost of the matrix-vector multiplication, the inversion of the 
block diagonal matrix and performance on the GPU and CPU. 

\begin{table*}[!ht]
\renewcommand{\arraystretch}{1.3}
\centering
    \begin{tabular}{r!{\vrule width 1.25pt}c|c|c|c|c}
        Name            & Processors        & Cores & Speed   & GFLOPS & Watts \\ \noalign{\hrule height 1.25pt}
        Interlagos (XE6)     & 2 $\times$ AMD 6272      & $4\times8=32$    & 2.1 GHz & {\raise.17ex\hbox{$\scriptstyle\sim$}}295    & 230  \\ \hline
        Maranello  & 2 $\times$ AMD 6168      & $4\times6=24$ & 1.9 GHz & {\raise.17ex\hbox{$\scriptstyle\sim$}}202    & 230  \\ \hline
        Fermi & 1 $\times$ NVIDIA C2050  & $14\times32=448$ & 1.2 GHz & {\raise.17ex\hbox{$\scriptstyle\sim$}}515    & 238  \\ \hline
        Kepler (XK7) & 1 $\times$ NVIDIA K20X & $14\times192=2688$ & 0.73 GHz & {\raise.17ex\hbox{$\scriptstyle\sim$}}1310    & 235 
    \end{tabular}\label{tab:systems}
\caption{Test systems for the performance benchmarks. The column 
\emph{Cores} shows the processor core configuration, the columns \emph{Speed} and \emph{GFLOPS} the chip clock speed and the 
theoretical peak performance in double precision of the cores and the column \emph{Watts} the 
power consumption in Watts.}   
\end{table*}
\begin{figure*}[!ht]
\centerline{
\subfloat[Patch size $16^{3}$ and block size of $8^{3}$.]
{\label{fig:numGauss16}\includegraphics[scale=1.0]{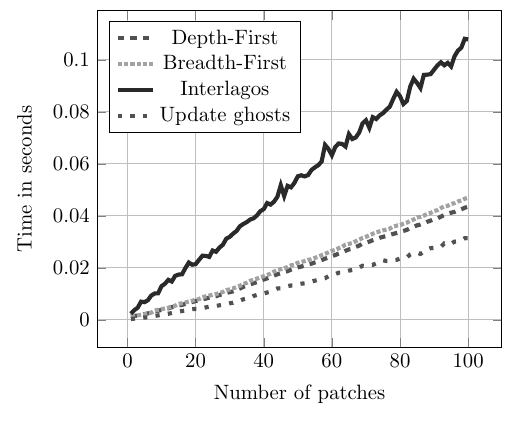}} \hfill
\subfloat[Patch size $64^{3}$ and block size of $8^{3}$.]
{\label{fig:numGauss64}\includegraphics[scale=1.0]{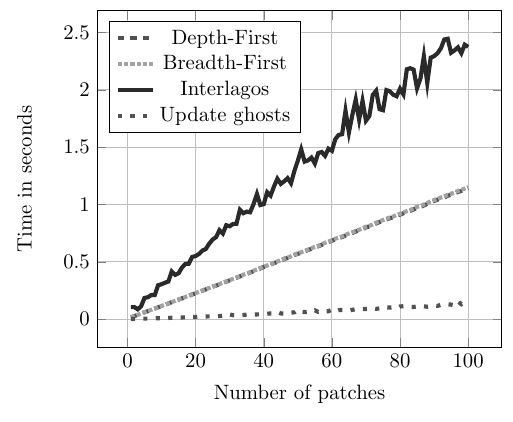}} 
}
\caption{Chaotic Gauss-Seidel relaxation on Kepler using different overlapping techniques.}
\label{fig:numPatches}
\end{figure*}

The results from the multicore benchmarks show that using the 
patch-parallelism, block-parallelism or two-level parallelism gives similar scalability,
especially when using block size of $ 8^3 $
though the patch-parallelism is the best for block size $ 4^3 $.
A maximum speedup around 18 is reached for the block Jacobi and 20 for chaotic block 
Gauss-Seidel using block size $8^{3}$.

This leads to an efficiency of about 80\%
on the 24-core machine. Choosing a block size of $4^{3}$ drops the maximal 
speedup for both algorithms to about 12 for block Jacobi and 
14 for chaotic block Gauss-Seidel. The patch-parallel version 
of the chaotic block Gauss-Seidel is in fact a standard block Gauss-Seidel 
relaxation without chaotic updates because the blocks are processed in serial. 

A larger block size leads to better 
scalability with respect to efficiency for both algorithms. The two-level 
parallelism scales as well as the patch-parallel version and allows the 
chaotic block update in the Gauss-Seidel scheme. 

From the speedup results we see that the block 
Jacobi algorithm gains more from processing the patches in parallel than 
processing the blocks in parallel, because the block-parallel version does not 
scale well with smaller block sizes. The patch-parallel or nested-parallel 
versions of Jacobi scale better because the solution $u^{k+1}$ is advanced 
for the patches in parallel. In the case of the chaotic block Gauss-Seidel 
the block parallel version scales better for smaller block sizes than the 
two-level parallelism. 
The stairstep pattern in the speedup of the two-level parallelism 
indicates a workload imbalance caused by adding block parallelism to the patch 
parallel version.

\subsection{Algorithm Performance}\label{sec:appperf}

\begin{figure*}[!t]
\centerline{
\begin{tabular}{cc}
\subfloat[Block Jacobi relaxation.]
{\label{fig:gcvJacobi}\includegraphics[scale=0.9]{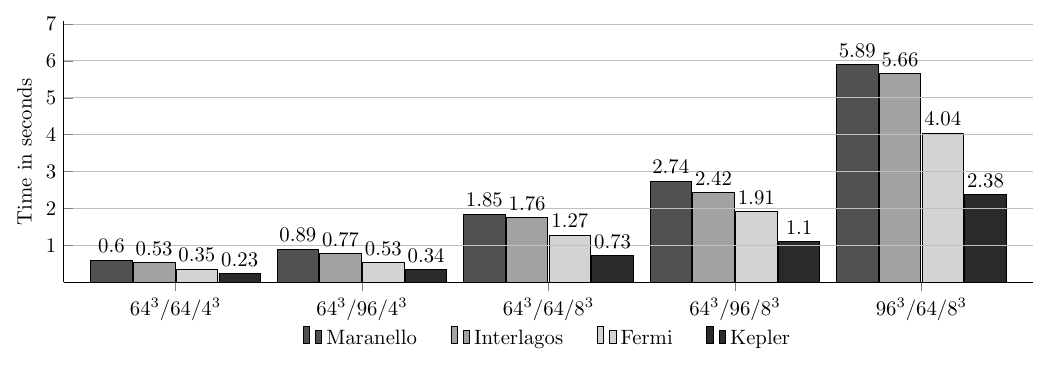}} \\
\subfloat[Chaotic block Gauss-Seidel relaxation.]
{\label{fig:gcvGauss}\includegraphics[scale=0.9]{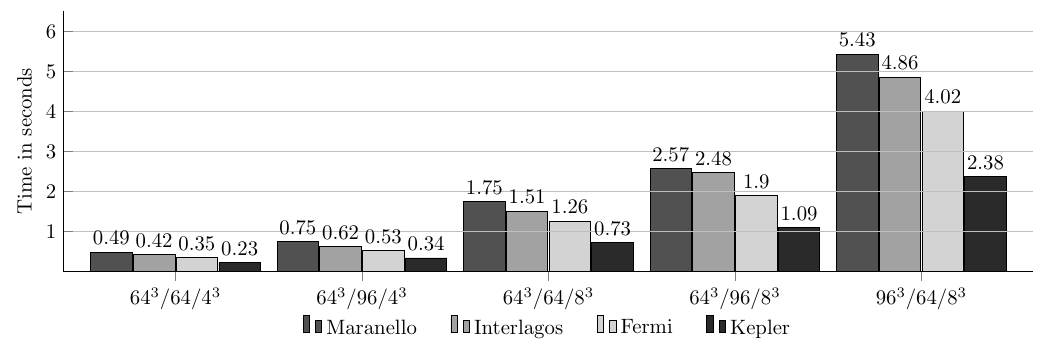}}
\end{tabular}
}
\caption{Time to smooth with chaotic block Gauss-Seidel and block Jacobi relaxations. 
The data is grouped by parameter set 
(patch size / number of patches / block size) for different systems.}
\label{fig:timeSmooth}
\end{figure*}

In this section we 
present measurements that have been taken on the systems listed in 
Table 1 and from now we reference the system by its name. 
One system named `Maranello' consists of two AMD Opteron 6168 CPUs and one 
NVIDIA TESLA C2050 (Fermi) \cite{NVIDIA:C2050:BRIEF}. 
The other system named `Interlagos' is 
a CRAY XE6 \cite{CRAY:XE6} node equipped with two AMD Opteron 6272 CPUs 
and a CRAY XK7 node \cite{CRAY:XK6} provides the benchmarks for the 
NVIDIA TESLA K20X (Kepler) GPU  \cite{NVIDIA:KEPLER}.
Floating-Point Units (FPUs) on the NVIDIA Fermi GPU are
arranged as a set of 32 in each Streaming-Multiprocessor (SM),
while on the NVIDIA Kepler GPU,
each Streaming-Multiprocessor (SMX) comprises of
192 single-precision CUDA cores, 64 double-precision units, 32 special function units, and 32 load/store units.
On the CPUs the arrangements of FPUs 
are different as for Maranello each die consists of 6 FPUs and for 
Interlagos each die has 4 FPUs but 8 integer units. The different FPU arrangements 
reflect one difference between the two AMD architectures. 
All benchmarks in this section 
exclude the time to compute the inverses.
Timings to compute the inverses are presented in Section \ref{sec:binpact}. 

The GPU has the option to overlap as described in Section \ref{sec:implementCuda}. 
A benchmark is used to explore the performance of the different overlapping 
techniques to estimate which method should achieve the best performance 
for different patch sizes. 
Figure \ref{fig:numPatches} shows the wall time for 3 smoothing steps 
varying numbers of patches and the patch size on the Kepler GPU
by exploiting kernel overlapping (breadth-first) 
or kernel transfer overlapping (depth-first). Figure \ref{fig:numPatches} also 
shows the wall time for the Interlagos system using the two-level parallelism 
for comparison. 

The depth-first overlapping strategy is slightly better than the breadth-first approach
for patch size of $ 16^3 $, and they give almost identical timings for patch sizes $ 64^3 $.
Moreover, they both outperform the CPU timings taken on Interlagos
by a factor around 2.5. Note that all these timings include
updating the ghost cells among the patches.
The bottom line in Figure \ref{fig:numPatches}a and \ref{fig:numPatches}b is the time
required to update the ghost cells.
When the patch size is $ 16^3 $, updating the ghost cells takes
about 70\% of the GPU time and 25\% of the CPU time,
and while the patch size is $ 64^3 $, it consumes around 15\%
of the GPU time and 5\% of the CPU time.
This indicates that using patch size of $ 64^3 $ or larger may be more preferable
in AMR applications.

For the benchmarks the focus is on patch sizes of at least 
$64^{3}$ so that the timings are taken with depth-first asynchronous 
memory transfer from and to the GPU in each smoothing step, while the CPU 
implementation does not need to execute memory transfers after each step.
Timings for the CPU code are 
taken from the two-level nested version. In Figure \ref{fig:timeSmooth} 
the wall time, smoothing selected parameter sets is shown for the block 
Jacobi and for chaotic block Gauss-Seidel relaxations. 

Three smoothing steps are executed on varying parameter sets. Each 
parameter set is  
defined by patch size, number of patches and block size. Sets with 
patch sizes of $64^{3}$ and $96^{3}$ and block sizes of $4^{3}$ and $8^{3}$ 
are benchmarked. 
The Interlagos outperforms the Maranello because of its higher peak performance. 

Both GPUs perform better than the multicore systems. The Kepler is 
faster then the Fermi because of the new architecture and higher peak performance.
The wall time using the parameter set ($64^{3}/96/8^{3}$) for both algorithms 
on the Kepler is about $2.2\times$ faster than the multicore time on the 32-core Interlagos machine.

\begin{figure*}[!t]
\centerline{
\subfloat[Block Jacobi relaxation.]
{\label{fig:inhomoJacobi}\includegraphics[scale=1.0]{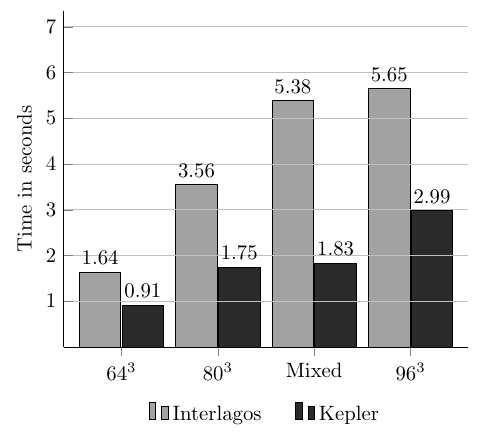}} \hfill
\subfloat[Chaotic block Gauss-Seidel relaxation.]
{\label{fig:inhomoGauss}\includegraphics[scale=1.0]{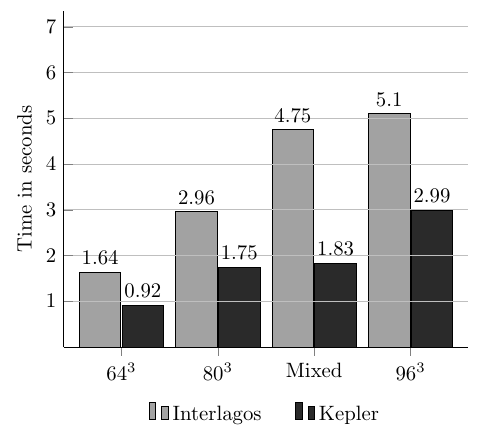}}
}
\caption{Smoothing sets of patches with the same and with mixed sizes.}
\label{fig:inhomo}
\end{figure*}

\begin{table}[!htbp]
\renewcommand{\arraystretch}{1.3}
\centering
    \begin{tabular}{c|c|c|c}
        Patch size & \# & \#Cells & Block size\\ \hline
        $64 \times 64 \times 64$ & 80 & 62,914,560 & $8\times8\times8$ \\ \hline
        $80 \times 80 \times 80$ & 80 & 122,800,000 & $8\times8\times8$ \\ \hline
        Mixed & 80 & 130,252,800 & $8\times8\times8$ \\ \hline
        $96 \times 96 \times 96$ & 80 & 212,336,640 & $8\times8\times8$
    \end{tabular}\label{tab:mixedPatches}
\caption{Patch sets for smoothing fixed number of 80 patches using sets 
of the same and mixed sizes.}   
\end{table}

The selected application parameter sets use the same patch sizes which are characteristic of one class of AMR methods.
However, for another class of AMR applications the patch size can vary.
We benchmark the performance of the block smoothers using different patch sizes to better reflect the practical  
performance for this latter class of AMR methods. For estimating the practical performance,  
a benchmark executing the smoothing algorithms on patch sets with the same patch 
sizes and sets with different sizes as listed in Table 2. 
All sets consist of 80 patches and the mixed patch set is divided into subsets 
of 16 patches with $64^{3}$, $72^{3}$, $80^{3}$, $88^{3}$ and $96^{3}$.
The ratio of cells between $64^{3}$ and $80^{3}$ is 1.89 and the ratio of cells  
between $64^{3}$ and the mixed set is 2.07. The ratio of cells between 
$80^{3}$ to $96^{3}$ is 1.72 and the ratio of cells between mixed and $96^{3}$ is 1.63.

Figure \ref{fig:inhomo} shows the wall time for
smoothing patches of the same and mixed sizes using block Jacobi or 
chaotic block Gauss-Seidel on the fastest GPU and the CPU system.
The time for smoothing the patches on the GPU increases roughly proportionally 
to the number of cells independent of the mixed patch sizes for both 
algorithms. 
This behavior is expected because no synchronization is needed at 
the end of a processed patch and the memory transfer overlaps with the 
kernel execution. Essentially the GPU just batch processes block after block 
without stalling.
However, the behavior is not the same in the CPU implementation. In the 
CPU implementation the wall time increases proportionally to the patch size for 
constant-size patches. For the mixed patch size set the 
wall time is closer to the time that is needed to smooth the largest  
patches. The reason for the poor performance of the mixed set is that the  
OpenMP threads looping over the blocks must \emph{join} after the patches are 
processed, and end up waiting for the largest patches.

Load balancing in the case of mixed 
patch sizes is an issue that depends on the ordering of the patches. In reality,  
the ordering of patches is random and in the worst case one thread gets all 
the larger patches. Using a naive approach by partitioning the patches without 
respect to the size has a potential for imbalance. The computation of an optimal 
partitioning is a variation of the bin packing problem which is known to be  
NP-hard. We 
can control the ordering of patches in the benchmark and if the patches are 
partitioned perfectly (each thread has the same amount of work) then the 
wall time for the mixed set drops to 4.37 seconds for block Jacobi and 
3.96 seconds for chaotic block Gauss-Seidel.

Smoothing 
using mixed patch sizes accelerates the GPU wall time more than $2.9\times$ 
over the CPU wall time in the case of block Jacobi relaxation and 
chaotic block Gauss-Seidel when the patches are imbalanced. 
A balanced patch partitioning 
improves the wall time on the CPU for the mixed set, but to be fair the 
additional cost for partitioning must be taken into consideration because 
it is not necessary on the GPU.

\begin{figure}[h]
\centering
\includegraphics[scale=0.9]{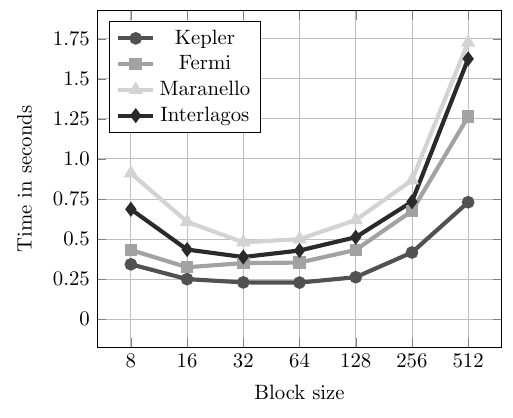}
\caption{Kernel performance for chaotic block Gauss-Seidel using different 
block sizes with patch size $64^{3}$.}
\label{fig:blockKernels}
\end{figure}

But as discussed in Section \ref{sec:blockDomain},
another important factor contributing to the wall time is the block size.
The number of floating-point operations for the stencil is constant 
for a thread in a thread block regardless to the block size, so   
the performance is dominated by the matrix-vector multiplication of 
the block inverse with the residual.
Figure \ref{fig:blockKernels} shows the wall time for smoothing 3 iterations 
on 64 patches with a patch size 
of $64^{3}$ using different block sizes for the chaotic block Gauss-Seidel method.

The GPU outperforms the CPU for all cases considered.
For a block size of $8^{3}$ more than 67 million 
cells per second can be smoothed on the Kepler and around 31 million cells per 
second on the Interlagos.  

\subsection{Block inversion}\label{sec:binpact}

Our block-relaxation algorithms require block inversions of  
diagonal blocks from the matrix that corresponds to the discretization 
over the AMR domain. In the case 
of constant-coefficient stencils the computation of the 
inverse must be performed only once if the patch size is a multiple of the block size. 
If the patch size is not a multiple of the block size then two possibilities 
exist. Either the blocks that do not fit at the end of the patch overlap with 
the previous ones, which can potentially change the convergence behavior, even resulting in  
divergence of the smoothing algorithm for chaotic updates,
or all inverses for these cases must be precomputed. 

\begin{figure}[h]
\centerline{
\includegraphics[scale=1.0]{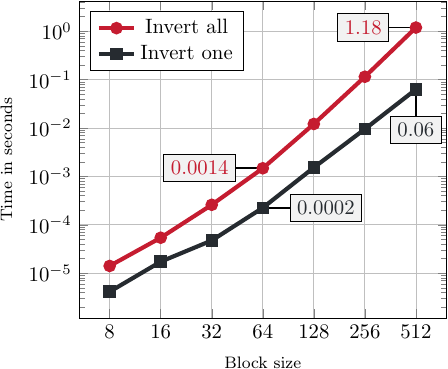}
}
\caption{Total time to compute the inverse of either one matrix or all block matrices 
for different block sizes.}
\label{fig:inverse}
\end{figure}

Computing all inverses 
is a robust approach that needs more memory and more computation time upfront. 
A total of $b_{n}=b_{x}b_{y}b_{z}$ inverses 
are potentially needed for a 
block of size $b_{x} \times b_{y} \times b_{z}$. To estimate the inverse 
computation overhead a benchmark computes either one inverse for a block 
or all inverses for a particular block size as shown in Figure \ref{fig:inverse}. 

The inversion is performed using a sequential BLAS library on the Interlagos system. 
Computing the inverses can be done in parallel but the cost of 
computing the inverses in serial is negligible in both cases. For example, computing 
the inverse of the block size $8^{3}$ takes 0.06 seconds on Interlagos, 
or about 4\% of the wall time of 1.51 seconds needed for one simulation 
iteration smoothing three steps with 64 patches of size $64^{3}$ using 
chaotic block Gauss-Seidel on the same system. 
In the case of multiple inverses this fraction increases but is also negligible 
because the inverses have to be computed only once for 
the whole simulation in the case of constant-coefficient stencils and therefore 
the time for precomputing the inverses is amortized over a few simulation steps. For example, in our
typical AMR simulations the inverses can be reused thousands of times.

\section{Conclusions and Future Work}

We implemented our block smoothing algorithm 
for structured adaptively refined meshes using block Jacobi and chaotic block 
Gauss-Seidel relaxations for modern GPUs and multicore CPUs. Our multicore 
implementation scales on a 24-core machine with an efficiency of 80\%, achieving  
a speedup around 20 by exploring multiple parallel strategies. The 
GPU version is about $2.2\times$ faster than our best multicore system using 
patches of the same sizes. In practical benchmarks with mixed patch sizes we 
can observe about $2.9\times$ improvement using a GPU.
The peak performance of Kepler is about $4.4\times$ the peak of the newest 
CRAY XE6 compute node. An overall wall time acceleration from 
$2.2\times$-$2.9\times$ 
on Kepler is a good result but there is still room to 
improve both implementations.

The way asynchronous memory transfer on the GPU will be issued does not
have an impact on the Kepler GPU.
Comparing the block Jacobi with the chaotic block Gauss-Seidel shows that 
both algorithms scale in the same fashion and the chaotic block Gauss-Seidel 
algorithm gives a better wall time. The chaotic block Gauss-Seidel 
gives better results in terms of convergence and relative error for the 
random initial guess for almost all tested block sizes. Increasing the block size further
improves the convergence factor for both algorithms.

In addition we presented strategies for the parallelization of the block 
smoothing algorithm in the context of adaptively refined meshes. Different 
abstract mapping strategies for the GPU were also presented and can be used 
independently from the smoothing in future structured AMR-based algorithms.
The implemented OpenMP versions use single-level and two-level parallelism 
approaches to realize these mappings.

Our future investigation is to optimize our block-relaxation 
implementations for 
the next hybrid supercomputer architecture by exploring more GPU parallelism 
like tiling the blocks or other patch mapping strategies. Moreover a hybrid 
implementation by smoothing patches on the GPU and CPU is necessary because 
the memory transfer and kernel execution on the GPU is asynchronous so that 
the CPU is idle during the processing on the GPU. This wasted compute 
power can be used in future implementations.
We also plan to use the implemented block relaxation algorithms
in multilevel preconditioners to solve complicated multi-physics problems
on distributed systems.

\section*{Acknowledgments}
This research was conducted in part under the auspices of the Office of 
Advanced Scientific Computing Research,
Office of Science, U.S. Department of Energy under Contract No. 
DE-AC05-00OR22725 with UT-Battelle, LLC.
This research used resources of the Leadership Computing Facility at Oak 
Ridge National Laboratory, which is supported
by the Office of Science of the U.S. Department of Energy under Contract 
No. DE-AC05-00OR22725 with
UT-Battelle, LLC. Accordingly, the U.S. Government retains a 
non-exclusive, royalty-free license to publish or reproduce
the published form of this contribution, or allow others to do so, for 
U.S. Government purposes.

Mark Berrill acknowledges support from the Eugene P. Wigner Fellowship at Oak 
Ridge National Laboratory, managed by UT-Battelle, LLC, for the U.S. Department 
of Energy under Contract DE-AC05-00OR22725.

The authors would like to thank Rebecca Hartman-Baker from iVEC for the very 
useful additions and corrections to this paper, James Schwarzmeier from 
CRAY Inc. for providing access to the CRAY XE6 system and Carl Ponder from 
NVIDIA Corporation for the useful discussions about CUDA.

\bibliographystyle{elsarticle-num}

\end{document}